\documentclass[12pt]{article}

\usepackage{graphicx}

\usepackage{iopams} 

\usepackage{bibentry}

\usepackage{times}

\newenvironment{sciabstract}{%
\begin{quote} \bf}
{\end{quote}}

\title{Electromechanical enhancement of live jellyfish for ocean exploration}

\author
{Simon R. Anuszczyk$^{1}$ and John O. Dabiri$^{1,2\ast}$\\
\\
\normalsize{$^{1}$Graduate Aerospace Laboratories, California Institute of Technology, Pasadena, CA, USA.}\\
\normalsize{$^{2}$Mechanical and Civil Engineering, California Institute of Technology, Pasadena, CA, USA.}\\
\\
\normalsize{$^\ast$To whom correspondence should be addressed; E-mail:  jodabiri@caltech.edu.}
}

\date{}

\begin{document} 


\baselineskip24pt


\maketitle


\begin{sciabstract}
The vast majority of the ocean's volume remains unexplored, in part because of limitations on the vertical range and measurement duration of existing robotic platforms. In light of the accelerating rate of climate change impacts on the physics and biogeochemistry of the ocean, the need for new tools that can measure more of the ocean on faster timescales is becoming pressing. Robotic platforms inspired or enabled by aquatic organisms have the potential to augment conventional technologies for ocean exploration. Recent work demonstrated the feasibility of directly stimulating the muscle tissue of live jellyfish via implanted microelectronics. We present a biohybrid robotic jellyfish that leverages this external electrical swimming control, while also using a 3D printed passive mechanical attachment to streamline the jellyfish shape, increase swimming performance, and significantly enhance payload capacity.  
A six-meter-tall, 13,600-liter saltwater facility was constructed to enable testing of the vertical swimming capabilities of the biohybrid robotic jellyfish over distances exceeding 35 body diameters. We found that the combination of external swimming control and the addition of the mechanical forebody resulted in an increase in swimming speeds to 4.5 times natural jellyfish locomotion. Moreover, the biohybrid jellyfish were capable of carrying a payload volume up to 105\% of the jellyfish body volume. The added payload decreased the intracycle acceleration of the biohybrid robots relative to natural jellyfish, which could also facilitate more precise measurements by onboard sensors that depend on consistent platform motion. While many robotic exploration tools are limited by cost, energy expenditure, and varying oceanic environmental conditions, this platform is inexpensive, highly efficient, and benefits from the widespread natural habitats of jellyfish.
The demonstrated performance of these biohybrid robots suggests an opportunity to expand the set of robotic tools for comprehensive monitoring of the changing ocean.
\end{sciabstract}


\section{\uppercase{Introduction}}
The ocean contains the majority of the habitable volume on earth, and many of its habitats and species remain largely unexplored \cite{ramirez2010deep}. The acceleration of climate change is leading to increasing ocean warming, acidification, and deoxygenation and posing a danger to biodiversity and associated ocean resources \cite{levin2015deep}. Recognizing these impacts, the United Nations has declared this the 'United Nations Decade of Ocean Science for Sustainable Development' in an effort to galvanize ocean exploration \cite{ryabinin2019decade}. Less than 80\% of the ocean has been explored, with basic properties such as depth known for only 18\% of the ocean at a resolution of 1 km \cite{wolfl2019seafloor}. These knowledge gaps motivate more comprehensive ocean exploration at timescales relevant to climate change \cite{benway2019ocean}. 

A variety of robotic technologies for ocean exploration have resulted in significant advances in ocean science \cite{yuh2011applications,zereik2018challenges}. For example, remote sensing platforms such as uncrewed aerial vehicles and satellites have been used to produce high resolution maps of sensitive marine habitats \cite{ventura2018mapping} and to study important physical characteristics such as salinity \cite{reul2014sea} and temperature \cite{parker1995marine}. However, these technologies are limited in the achievable depth of observations by water turbidity and solar illumination \cite{kavanaugh2021satellite}. Deep sea tools including gliders \cite{rudnick2016ocean}, seafloor mounted buoys \cite{favali2006seafloor}, and autonomous underwater vehicles \cite{sahoo2019advancements} can explore to significant depths but with mission duration often limited by battery capacity. The sheer size of the ocean, comprising more than 1.3 billion cubic kilometers \cite{charette2010volume}, challenges the cost-effective scaling of these technologies for comprehensive ocean coverage \cite{zereik2018challenges}.

Robotic vehicles wholly comprised of mechanical components have increasingly been inspired by the demonstrated performance of aquatic animals, with the goal of improving robotic vehicle propulsive efficiency and range. For example, an acoustically controlled soft robotic fish was developed to observe aquatic animal behaviour with minimal disruption  \cite{katzschmann2018exploration}. However, this robot required an accompanying human diver, thereby limiting operating depths to regions that are safe for human diving. A benthic walking robot demonstrated locomotion along the seafloor and geological sediment-sampling using a shovel \cite{picardi2020bioinspired}. While valuable for benthic exploration, this and similar robots are limited to the seafloor due to their inability to swim. A soft snailfish-inspired robot demonstrated swimming locomotion at a depth of 10,900 m  \cite{li2021self}. This work presented a novel method for electronics hardening by distributing the electronics in a polymeric matrix, but power consumption limited swimming to 45 minutes. Each of the aforementioned examples demonstrates the emerging capabilities of robotic mimics of aquatic organisms; however, the performance envelope of live aquatic animals in the ocean has not yet been realized by these approaches.

A different strategy, biologging, embeds sensor tags on live animals such as pinnipeds \cite{carter2016navigating}, sea turtles \cite{cuevas2008post}, and jellyfish \cite{mooney2015itag,fossette2016tag} and records measurements of the aquatic environment as the animals locomote naturally through the ocean. Biologgers can be used to study ocean characteristics such as salinity, temperature, and oxygen distribution using conductivity-temperature-depth instruments, oxygen probes, hydrophones, and other sensors \cite{chung2021review}. However, the timing and locations of the measurements are uncontrolled and limited to the path of the tagged animal. 

Biohybrid robots have the potential to combine the control of fully mechanical robotic platforms and the swimming performance envelope of live aquatic animals to achieve the goal of comprehensive ocean exploration. Proof-of-concept of biohyrid robotic control has been achieved in a variety of organisms, including beetles \cite{sato2009remote}, cockroaches \cite{sanchez2015locomotion}, hawkmoths \cite{bozkurt2009balloon}, and tissue-engineered rays \cite{park2016phototactic} and jellyfish \cite{nawroth2012tissue}. Some biohybrid robots have also made use of biofuel cells to power onboard electronics \cite{schwefel2014wireless, shoji2016autonomous} and sensing for insect swarms \cite{romano2023beehive}. By utilizing biological locomotion, biohybrid robotics can potentially capitalize on the efficiency and adaptability of existing biological organisms. Biohybrid robotics have significant potential but so far have largely been confined to proof-of-concept and have not yet been deployed as platforms for science. 

Here, we investigate jellyfish as a platform for biohybrid robotic sensors that can access the full ocean, including its deepest regions. Jellyfish are naturally occurring in a diverse range of oceanic environments, including tropical \cite{purcell2012jellyfish} and polar regions \cite{purcell2010distribution}, and spanning from the surface to depths exceeding 10,800 meters \cite{gallo2015submersible}. They are also tolerant to hypoxic conditions below 1 mg O\textsubscript{2} per liter \cite{purcell2007environmental}, in contrast to fish who avoid levels below 2 to 3 mg O\textsubscript{2} per liter. In terms of their locomotion, jellyfish are the most energy efficient of all known metazoans, with a cost of transport 3.5-fold lower than even efficiently swimming fish such as salmon \cite{gemmell2013passive}. 

Jellyfish have previously been used as biologgers to investigate their ecology and behaviour \cite{mooney2015itag,fossette2016tag}. Previous work has also demonstrated that the swimming of live jellyfish can be controlled by external electrical stimulation from an implanted microelectronic device \cite{xu2020low}. When swimming contractions were stimulated at approximately 0.6 Hz, the jellyfish were observed to swim up to 2.8 times baseline swimming speeds without the swim controller. These increased speeds required only a twofold increase in energy expenditure, implying a concomitant increase in propulsive efficiency during external swimming control. The aforementioned adaptability to a wide range of ocean regions and high energy efficiency make jellyfish excellent candidates for a biohybrid robot.

In consideration of the ethics of the aforementioned robotic manipulations, the authors of that work collaborated with bioethicists to establish a framework for responsible implementation of these techniques \cite{xu2020ethics, Xu2024transanimalism}. This framework suggests consideration of the 4 "Rs" of reduction, replacement, refinement, and reproducibility which we have utilized in this study. It is important to note that the jellyfish genus that is the focus of this work, \textit{Aurelia}, lacks a central nervous system and nociceptors, and therefore lack the capacity to sense pain \cite{smith2009nociceptors}. While they have a known mucosal stress response, this was not observed to occur in any prior experiments using external swimming control.

Although previous work demonstrated the feasibility of external control of jellyfish swimming frequency and quantified associated performance gains, key gaps remain in development of the biohybrid jellyfish robot as a research platform for ocean science. First, the prior work did not explicitly incorporate a payload capacity, which is essential for the carrying of onboard measurement sensors. Second, assessment of the swimming performance was limited to tests either in a 2-meter-tall laboratory water tank \cite{xu2020low} or a coastal marine environment \cite{xu2020field}. Swimming performance over longer vertical distances consistent with the envisioned deep ocean measurements has not been characterized. Finally, while natural jellyfish boast a low cost of transport \cite{gemmell2013passive}, this performance is achieved despite a body shape that is not streamlined.

We investigate for the first time the feasibility of simultaneous electrical and mechanical modification of live jellyfish as biohybrid robotic platforms. The mechanical modification comprises the addition of a passive forebody to streamline the natural jellyfish body shape while also facilitating a significant payload capacity. The performance of the electromechanically modified jellyfish is tested in a specially constructed six-meter-tall, 13,600-liter saltwater facility where the jellyfish can be observed swimming over distances up to 35 body diameters. These measurements, combined with an analytical model of the swimming dynamics, enable optimization of the design of the biohybrid jellyfish robots.

\section{\uppercase{Methods and Materials}}

\subsection{Animal husbandry}
We conducted long-distance swimming experiments with $N=3$ jellyfish of diameters $d_1=16.0 \pm0.5$ cm, $d_2=16.5 \pm0.5$ cm, and $d_3=18.0 \pm0.5$ cm and flow visualization experiments with 1 jellyfish of diameter $d_4=19.0 \pm0.5$ cm. Utilizing jellyfish of varying geometries allowed us to draw broader conclusions about the applicability of this study to a range of animals. \textit{Aurelia aurita} were obtained from Cabrillo Marine Aquarium (San Pedro CA, USA) and housed in a 453-liter psuedokreisel tank (Jelliquarium 360, Midwater Systems, Thousand Oaks, CA, USA). The tank was filled with artificial seawater made from sea salt (Instant Ocean Sea Salt, Spectrum Brands,
Blacksburg, VA, USA) and deionized water balanced at 35 parts per thousand (PPT) and kept at 21$^{\circ}$C. The jellyfish were hand fed twice daily with live \textit{Artemia franciscana} brine shrimp (Hatching Shell-Free Brine Shrimp Eggs E-Z Egg, Brine Shrimp Direct, Ogden, UT, USA). Brine shrimp were hatched every other day in aerated hatchery cones of artificial seawater heated to 28$^{\circ}$C for 24 hours, then transferred to beakers at 21$^{\circ}$C. Brine shrimp were fed PHYTO-Feast (Reef Nutrition, Reed Mariculture Inc., Campbell, CA) and enriched with SELCO (Self-Emulsifying Lipid Concentrate, Brine Shrimp Direct, Ogden, UT, USA). The animals’ care was in accordance with institutional guidelines.

\subsection{System architecture}
Figure 1(a) shows the biohybrid robot electronics schematic and all components. The aspect ratio of the forebodies is defined as $AR=H/D$, where $H$ is the forebody height and $D$ is the maximum forebody diameter. The system comprises the jellyfish bell (1), the 3D printed $2.50 \pm 0.03$ cm diameter electronics housing containing the swim controller and battery (2), the two electrodes embedded in the jellyfish muscle tissue (3), a wooden pin attaching the electronics to the jellyfish (4), a 3D printed hemi-ellipsoid with $AR_1=0.31 \pm 0.01$ (5), a hemi-ellipsoid with $AR_2=0.66 \pm 0.01$ (6), and a hemi-ellipsoid with $AR_3=1.00 \pm 0.01$ (7). Figure 1(b) is a sample image of a jellyfish biohybrid robot with attached $AR_3$ forebody swimming in a lab saltwater tank.

\begin{figure}[!h]
\includegraphics[width=1\textwidth]{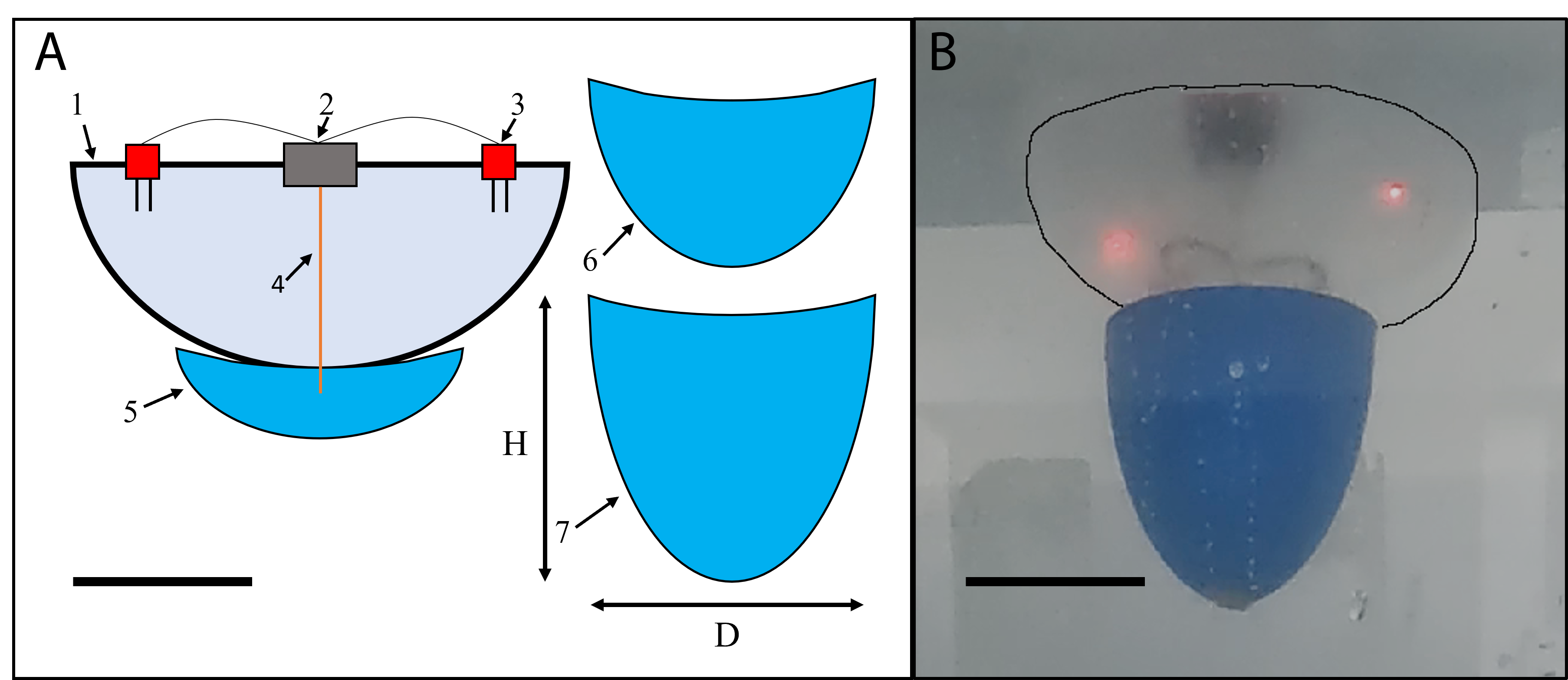}
\textbf{Fig 1. Hardware schematic.} \\
(\textbf{A}) Biohybrid robot schematic with 5 cm scale bar (horizontal black line). H and D are the height and diameter used to calculate the aspect ratio, $AR=H/D$. 1. Jellyfish bell. 2. Housing with $2.50 \pm 0.03$ cm diameter containing swim controller and battery. 3. Electrode embedded in muscle tissue. 4. Wooden pin attaching electronics to jellyfish. 5. Polylactic acid 3D printed forebody with $AR_1=0.31 \pm 0.01$. 6. Forebody with $AR_2=0.66 \pm 0.01$. 7. Forebody with $AR_3=1.00 \pm 0.01$. (\textbf{B}) Jellyfish biohybrid robot with attached hemi-ellipsoid with $AR_3$. This image has been digitally manipulated to remove reflections and outline the animal body in black.
\end{figure}

\subsection{Components and fabrication}
The electronics were based on previous work \cite{xu2020low} and consisted of a TinyLily processor (TinyCircuits, Akron, OH, USA) and a 15 mAH lithium-polymer battery cell (PGEB181118, PowerStream Technology Inc., Orem, UT, USA) enclosed in a custom $2.50 \pm 0.03$ cm diameter 3D printed housing printed on a resin printer for waterproofing at high pressure (Form 3, Formlabs, Somerville, Massachusetts). The housing utilized a male threaded base and female threaded cap compressing a face-sealed soft Buna-N O-ring with food-safe O-ring grease. Silver wire coated in perfluoroalkoxy was threaded through small holes in the housing and connected to platinum rods (A-M Systems, Sequim, WA, USA) embedded in the animal muscle tissue to create two electrodes. The holes were sealed with adhesive and and an LED was connected in series to each electrode to provide visual confirmation of stimulation as shown in figure 1(b).

Hemi-ellipsoid forebodies were designed with a density gradient such that the anterior portion, the bottom as shown in figure 1(b), was more dense than the posterior portion. Perturbations from vertical swimming, either due to endogenous swimming contractions by the animal or external fluid disturbances, were resisted by the passive restoring torque created by the position of the center of gravity below the center of buoyancy. Hence, the biohybrid robot was constrained to swim in the vertical direction, as envisioned for vertical ocean profiling. Concavity at the forebody-animal interface created a close fit against the exumbrellar bell surface. The forebodies were 3D printed out of PLA (Polylactic Acid) on an X1 Carbon printer (Bambulab, Shenzhen, China) and coated with 2-part epoxy resin for waterproofing. After drying, a hole was drilled into the forebody at the interface with the animal to create a press fit for the wooden rod as shown in figure 1(a) item 4. The forebodies were ballasted to maintain $0.01 \pm 0.01$ grams positive buoyancy in saltwater at 35 PPT. 

\subsection{Flow visualization experiments}
Flow visualization experiments to observe the effect of the forebody streamlining were conducted in a 1.2 m by 0.5 m by 0.5 m tank filled with artificial seawater at 35 PPT and seeded with 13 micron silver-coated hollow glass particles (SH400S20, Potters Industries, Carlstadt, NJ, USA). The biohybrid robot swam down through a laser sheet generated by a 671 nm continuous wave laser (5-Watt LRS0671 DPSS Laser System, Laserglow Technologies, North York, Ontario, Canada) shone through a 250 mm diameter condenser lens and a sheet forming optic. A high speed camera normal to the laser sheet recorded at 125 frames per second with a shutter speed of $1/200$ seconds, a resolution of 1024x1024 pixels (FASTCAM SA-Z, Photron USA Inc, San Diego, CA, USA), and a fixed focal length macro lens (Micro-NIKKOR 105 mm with a 36 mm extension tube, Nikon, Tokyo, Japan). The images were processed in MATLAB (Mathworks, Natick, MA, USA) by stacking every third frame for 50 frames. In each frame, the biohybrid robot was translated to the location in the previous frame such that the biohybrid robot was motionless. This manipulation transformed the video into the biohybrid robot frame of reference, and consequently, the observed particles traced out pathlines as in a long exposure image. This process was repeated for the biohybrid robot without the forebody and with the $AR_3$ forebody configuration to compare the streamlining effects.

\subsection{Long-distance swimming experiments}
A 6 m tall 13,600 liter saltwater facility was constructed for testing jellyfish biohybrid robot swimming performance. This 6 m by 1.6 m by 1.4 m, or 13.4 m\textsuperscript{3}, vertical tank was filled with artificial seawater balanced at 35 PPT and held at 21$^{\circ}$C (figure 2). The facility has a 151 liter per minute continuous filtration pump which turns over the entire system water volume every 90 minutes. This pump was turned off at least 2 hours before beginning experiments to allow residual currents to dissipate. The facility has three acrylic windows on each of three sides, spanning the width of the tank. A digital camera (GoPro HERO9 Black, GoPro Inc, San Mateo, CA) recording at a resolution of 3840x2160 pixels was used to record videos of the jellyfish biohybrid robot swimming trials. The camera was centered horizontally at a height of 1.8 m from the base of the tank and set back 2.1 m from the front. A 100 watt LED floodlight illuminated the tank from below through a porthole on the bottom of the tank.    

While salinity and temperature were carefully matched between the pseudokreisel and this facility, the jellyfish were also acclimated by slowly introducing water from this facility to the jellyfish for 30 minutes in a 5 gallon bucket. Each jellyfish was then transferred to a glass bowl and equipped with the electronics by inserting the wooden pin through the jellyfish stomach and embedding both electrodes into the muscle tissue on opposite sides of the bell margin. The jellyfish was transferred into the saltwater facility and the appropriate forebody was mounted to complete the biohybrid robot. A slack polymer string with 0.3 mm diameter was tied onto the electronics housing to facilitate recovery. The biohybrid robot was positioned horizontally in the center of the tank to avoid contacting the walls and released to begin swimming. Swimming data from the middle 2 m section of the tank was used in order to ensure the animal had reached a constant average swimming speed, and to avoid the hydrodynamic impact of the tank bottom. Post-processing of the measurements confirmed that the biohybrid robot reached a constant average swimming speed for all forebody configurations before entering the region of interest in the middle 2 m section of the tank. The sequence of forebody aspect ratio tests was randomized to minimize any systematic biases associated with a drift in animal swimming performance over the course of the swimming trials. Each forebody configuration was tested 3 times for a total of 12 trials per biohybrid robot. Between each trial, the biohybrid robot was retrieved using the polymer string and the tank was allowed to return to quiescence for 10 minutes before beginning the next trial. Experiments were conducted with all 3 jellyfish over 2 days.

\begin{figure}[!h]
\includegraphics[width=1\textwidth]{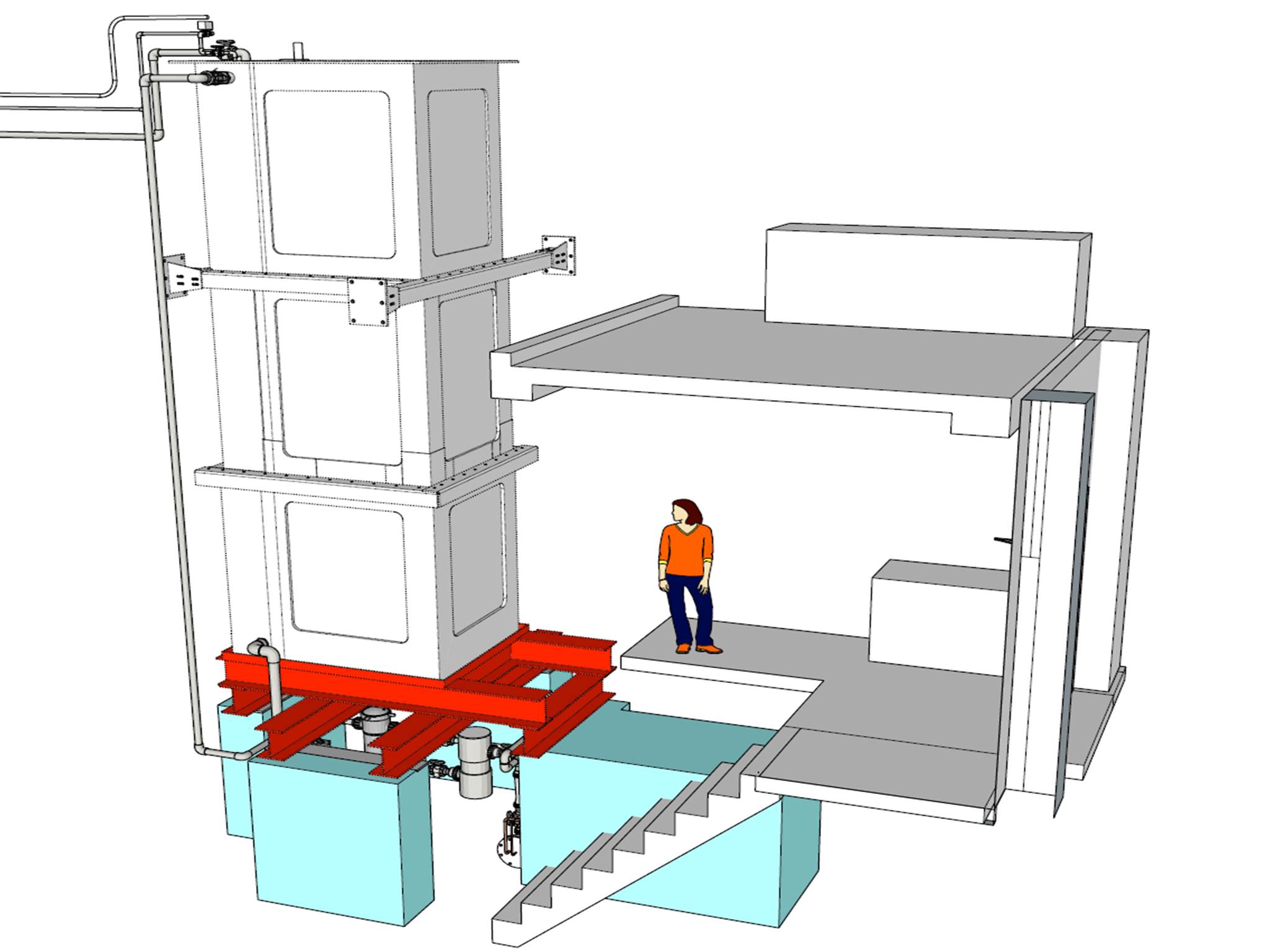}
\textbf{Fig 2. Saltwater vertical tank facility.} \\
The saltwater vertical tank facility used in the long-distance swimming experiments measured 6 m tall by 1.6 m by 1.4 m, or 13,600 liter, shown here with human for scale. Biohybrid jellyfish swimming trajectories were recorded with a digital camera placed at a height of 1.8 m from the base of the tank and horizontal distance of 2.1 m from the face of the tank.
\end{figure}

 Post-processing used a combination of a custom MATLAB tracking program and a commercial motion tracking software (Blender Foundation, Amsterdam, Netherlands) to track the jellyfish. The MATLAB tracking program assigned a tracking box around the biohybrid robot and used the centroid of the box to calculate biohybrid robot swimming speeds. The average swimming speed per trial was computed based on the slope of a linear regression of the measured swimming distance versus time. The best-fit line exhibited an R-square value of $0.9894 \pm 0.0141$, indicating the appropriateness of a linear regression. To determine instantaneous velocities during each trial, Blender was used to track the location of a point chosen near the centroid of the biohybrid robot. This tracking method better reproduced the intracycle speed changes as the jellyfish biohybrid robot contracted and relaxed but was not able to track the jellyfish without a forebody. Blender position data was imported into MATLAB where velocities were calculated as finite differences. Horizontal velocities were assumed as proxies for out of plane velocities and all 3 components were taken into account for all calculations. Intracycle speed changes were found with a program in MATLAB that identified peaks and valley in the speed trace and calculated the difference from a peak to the closest valley. Average speeds for both programs were the same. The steady-state swimming speed  for each trial was characterized by a drag coefficient $C_D$, corresponding to the measured terminal velocity with scaling coefficients $u=0.1417+0.0225\sqrt{1/C_D}$ where $u$ is the observed speed.

The forebody drag coefficients $C_D$, used in calculating the streamlining coefficient, $S=-log(C_D)$, were found by conducting drop tests of the forebodies. Forebodies of the same form factor as those used in our experiments but at a higher infill density were printed out of PLA such that they were negatively buoyant. These forebodies were attached to a slack polymer string with 0.3 mm diameter for ease of recovery, submerged at the top of the vertical tank, and released to sink at terminal velocity. These experiments were recorded using the same GoPro camera setup and post-processing methods as described above. The coefficients of drag $C_D$ were found using the drag equation \cite{batchelor1967introduction} which after accounting for buoyancy becomes 

\begin{equation}
C_D = \frac{2g(m_f-V_f\rho_w)}{\rho_wS_fu^2}
\end{equation}

where $g$ is the acceleration of gravity, $m_f$ is the mass of the forebody, $V_f$ is the volume of the forebody, $\rho_w$ is the density of saltwater, $S_f$ is the frontal surface area of the forebody impacting the flow, and $u$ is the terminal velocity at which the forebody falls. Each forebody was tested three times and the average velocity was used to calculate $C_{D,1}=0.93 \pm 0.05$, $C_{D,2}=0.52 \pm 0.03$, and $C_{D,3}=0.17 \pm 0.01$ where the error is the standard deviation of all three trials. The drag on the jellyfish without a forebody was approximated as a flat plate with $C_{D,0}=1.12$ \cite{hoerner1958fluid}.

\subsection{Kinematics Model}
We derived a first-principles physics-based kinematics model for our biohybrid robot by extending previous models \cite{daniel1983mechanics,mchenry2003ontogenetic,xu2020low} to predict performance of biohybrid robots with various payload and animal geometries. Starting with a force balance from Newton's second law

\begin{equation}
\sum{F} = m_R\frac{\partial u}{\partial t}
\end{equation}

where $\sum{F}$ is the sum of the forces, $m_R$ is the combined mass of the biohybrid robot, $u$ is the swimming speed, and $t$ is time. The thrust of the biohybrid robot $T$ is assumed to oppose drag $D$ and the unsteady acceleration reaction $Ac$

\begin{equation}
T-D-Ac = m_R\frac{\partial u}{\partial t}
\end{equation}

with the forces given as 

\begin{equation}
T=\frac{\rho_w}{A_{sub}}(\frac{dV_{R}}{dt})^2
\end{equation}

\begin{equation}
D=\frac{1}{2}C_D\rho_wS_Ru^2
\end{equation}

\begin{equation}
Ac=\alpha m_R\frac{\partial u}{\partial t}
\end{equation}

where $\rho_w=1.024$ g/cm$^3$ is the density of saltwater at 35 PPT and 21$^{\circ}$C, $A_{sub}$ is the subumbrellar opening area of the jellyfish, $V_R$ is the instantaneous volume of the biohybrid robot, $C_D$ is the coefficient of drag calculated above, $S_R$ is the instantaneous frontal surface area of the biohybrid robot, $\alpha=(h/r)^{1/4}$ is the added mass coefficient where $h$ is the bell height and $r$ is the bell radius. We define

\begin{equation}
V_R=V_f+V_0+\frac{dv}{dt}t
\end{equation}

\begin{equation}
S_R=S_f+S_0+\frac{3}{2h}\frac{dv}{dt}t
\end{equation}

where $V_f$ is the volume of the forebody, $S_f$ is the frontal surface area of the forebody, the relaxed volume of the jellyfish $V_0=2/3\pi hr^2$, the relaxed frontal surface area of the jellyfish $S_0=\pi r^2$, the change in volume $dv=V_{ex}V_0$ with a constant volume exchange of $V_{ex}$, change in time $dt=t_c$ during contraction and $dt=t_r$ during relaxation. Substituting these equations back into the force balance leads to 

\begin{equation}
\frac{\rho_w}{A_{sub}}(\frac{dV_{R}}{dt})^2=\frac{1}{2}C_D\rho_wS_Ru^2+(1+\alpha)m_R\frac{\partial u}{\partial t}
\end{equation}

which was solved for $u$ in MATLAB using the ODE45 function. Jellyfish geometry was measured to experimentally determine $h$ and $r$ for each jellyfish, and video analysis led to approximate contraction and relaxation times of $t_c=0.5$ and $t_r=1.5$ respectively.

\section{\uppercase{Results}}
We modified the shape of the exumbrellar surface of the natural jellyfish by attaching additively manufactured hemi-ellipsoids of various aspect ratios to the apex of the animals (see Methods). As in prior work \cite{xu2020low}, the jellyfish swim muscles, which are located on the subumbrellar surface, were stimulated with two electrodes to produce regular contractions at 0.5 Hz with a 3.7-volt, 10-ms square wave. The biohybrid robot is ballasted to create a slight positive buoyancy and a horizontal restoring torque constraining the animals to swimming in the vertical direction. This positive ballasting ensured that the observed downward swimming could be attributed only to the active animal swimming and not to the passive dynamics of the attached mechanical forebodies. 

\begin{figure}[!h]
\includegraphics[width=1\textwidth]{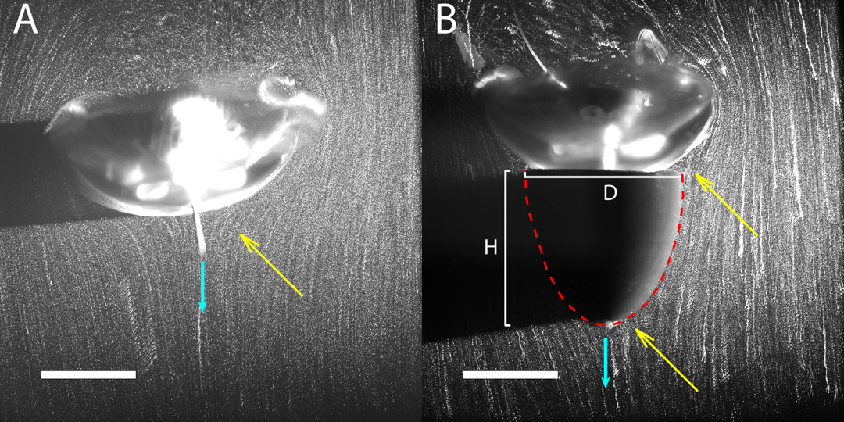}
\textbf{Figure 3. Comparison of flow streamlining with and without passive mechanical forebody.} \\
Jellyfish biohybrid robot with and without attached hemi-ellipsoid forebody. Cyan arrows indicate downward swimming direction of the animals. Yellow arrows highlight areas of flow diversion and demonstrate flow disturbance before (\textbf{A}) and after (\textbf{B}) addition of streamlined passive mechanical forebody (outlined in red dashed curve). White scale bar in each panel indicates 5 cm. In panel (B), $H$ is the forebody height and $D$ is the maximum forebody diameter. In both panels, the flow visualization was created by superimposing several consecutive images, each shifted spatially so that the jellyfish appears stationary while particles trace pathlines around the animal.
\end{figure}

\subsection{Flow visualization experiments}
We first studied the hydrodynamics of the biohybrid robot in a 1.2-m tall, 300 liter saltwater tank to investigate the hydrodynamic impacts of the mechanical forebodies on the local flow around the jellyfish. Time-lapse images of 13 micron, neutrally buoyant tracer particles showed that the addition of hemi-ellipsoids of increasing aspect ratios streamline the flow around the biohybrid robots (figure 3; see also Supplementary Movies 1 and 2). Specifically, comparison of the tracer particle pathlines illustrated the reduction in flow re-direction at the apex of the biohybrid robots relative to the natural jellyfish (yellow arrows). While the total lateral diversion of flow is the same in the two cases, i.e., directing flow around the maximum body width at the bell margin, the passive forebody achieves this diversion in a more gradual fashion. It was anticipated that the larger lateral diversion of flow at the apex of the natural jellyfish shape would lead to greater drag than on the streamlined forebodies.

\begin{figure}[!h]
\includegraphics[width=1\textwidth]{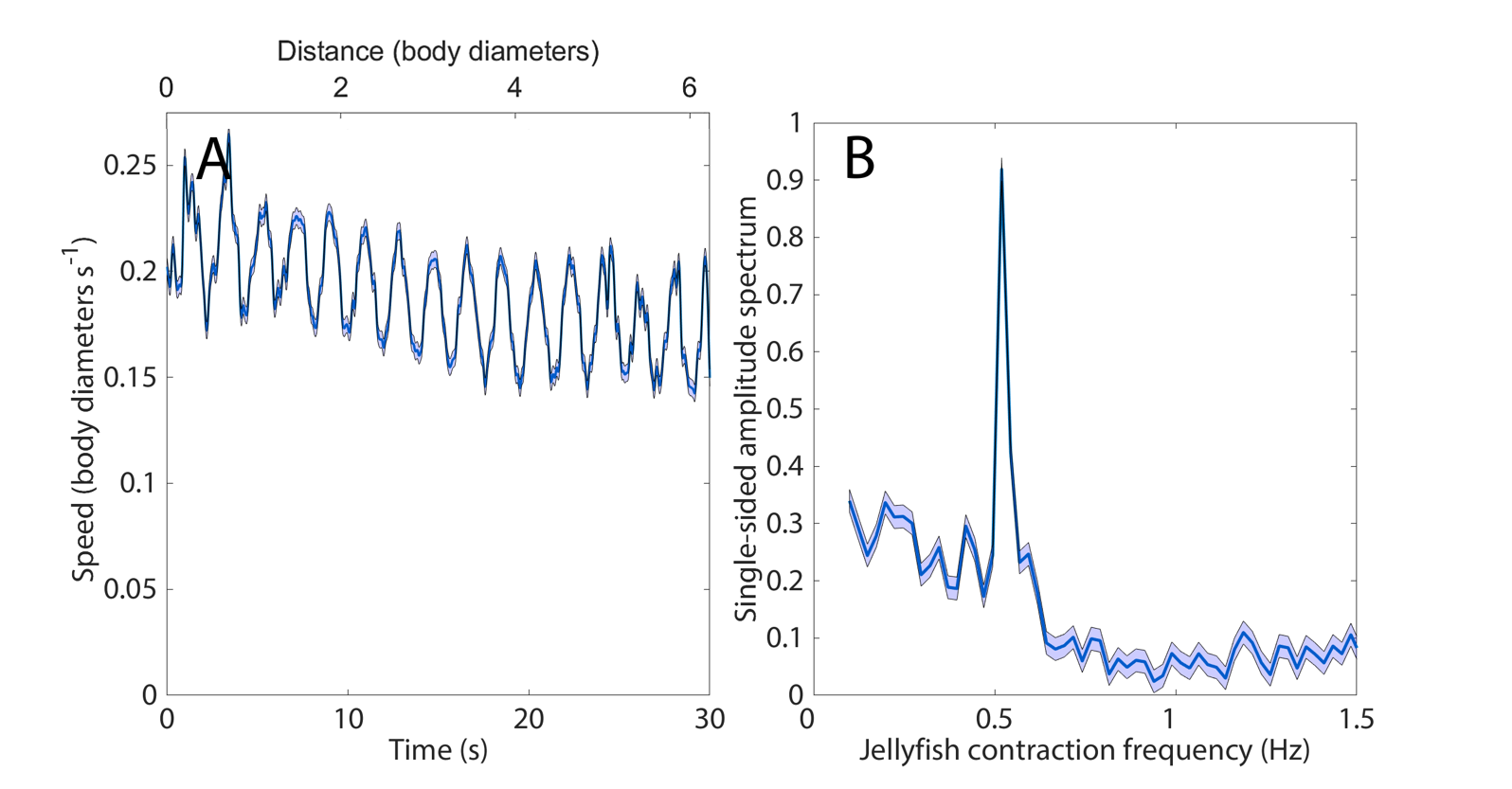}
\textbf{Figure 4. Example swimming speed measurements of biohybrid robot.} \\
(\textbf{A}) Contraction and relaxation cycles show example 30-second swimming speed trace (i.e., swimming distance of 6.2 body diameters) of biohybrid robot with 18 cm animal and attached forebody $AR_3$. (\textbf{B}) Frequency spectrum analysis confirms animal response to external stimulation at 0.5 Hz. Semi-transparent band represents a tracking error of $\pm 5$ pixels.
\end{figure}

\subsection{Long-distance swimming experiments}
To test the impact of the streamlining on swimming performance, the natural jellyfish and biohybrid robots were tested in a new 6 m tall, 13,600 liter saltwater facility built for free-swimming experiments (see Methods for facility description). Figure 4(a) shows a 30-second swimming speed trace (i.e., swimming distance of 6.2 body diameters) of the swimming biohybrid robot showing speed peaks during the jellyfish bell contraction phase and troughs during bell relaxation. The swimming speed oscillated with each contraction and relaxation, and eventually reached a steady state average swimming speed. A spectral analysis of the swimming speed trace shows that the animal responded to external electrical stimulation by contracting at the 0.5 Hz stimulation frequency (figure 4(b)). Figure 4(a-b) was made using the Blender tracking program.

\begin{figure}[!h]
\begin{center}
    \includegraphics[width=0.7\textwidth]{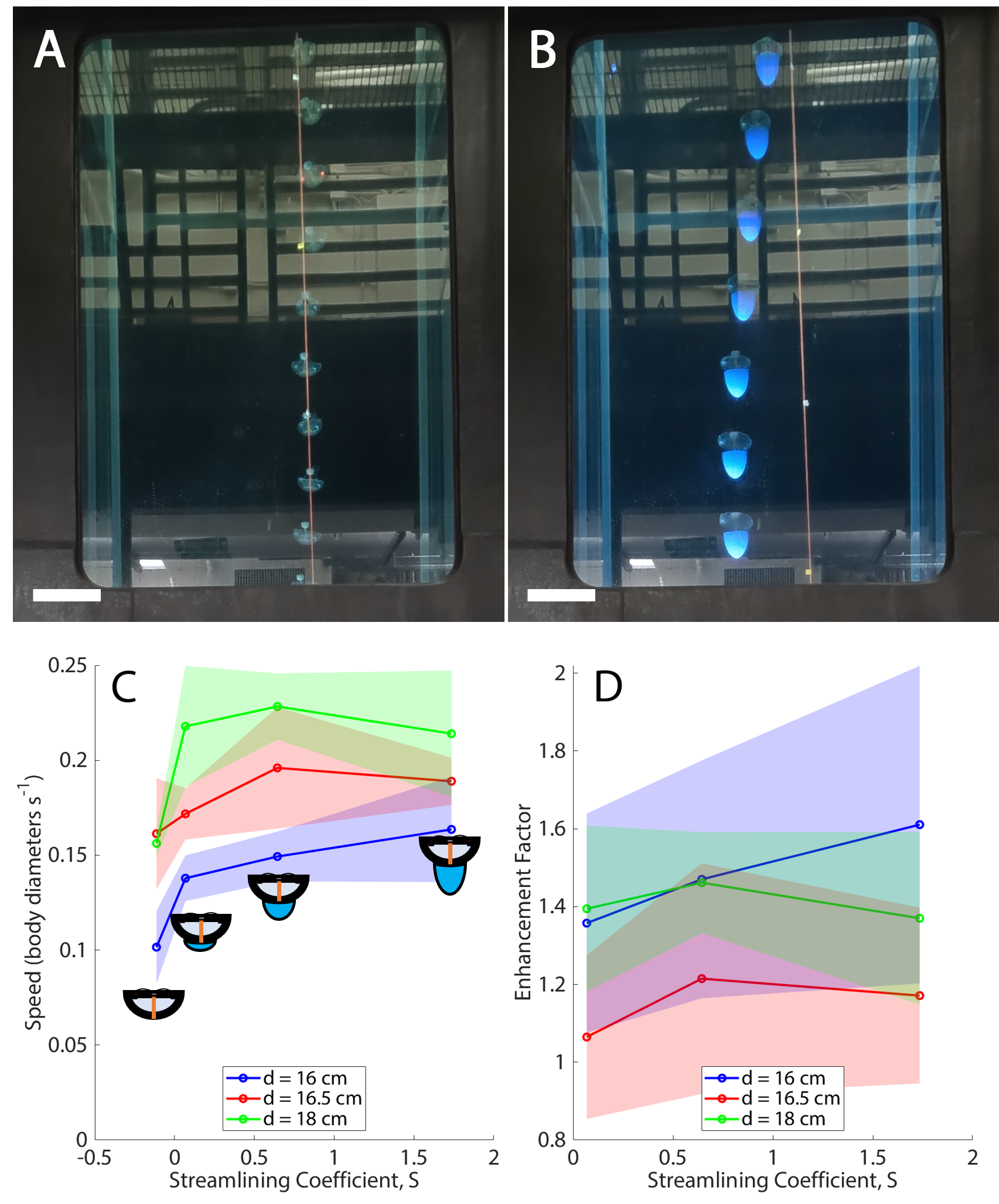} \\
\end{center}
\textbf{Figure 5. Measurements of swimming speed in six-meter vertical saltwater facility.} \\
Sample biohybrid robot trajectories at 6.67-s intervals for the biohybrid robot without mechanical forebody (\textbf{A}) and with mechanical forebody $AR_3$ (\textbf{B}). Larger spacing between robot images in panel (\textbf{B}) qualitatively illustrates faster swimming speed. Zoomed-in image shows middle one-third of height of facility. Orange dropline is marked every 1.5 m. White scale bar indicates 20 cm. (\textbf{C}) Normalized biohybrid robot speed increases with streamlining coefficient (S) demonstrating impact of streamlined forebodies. Each color represents one animal with diameters listed in the legend; transparent band represents $\pm$ 1 standard deviation, calculated based on the three trials at each point. (\textbf{D}) Enhancement factor compares swimming performance of biohybrid robot to same stimulated jellyfish without mechanical forebody. Maximum enhancement factor corresponds to 61\% increase in swimming speeds or a combined increase from electrical and mechanical enhancement to 4.5 times typical unstimulated jellyfish locomotion.
\end{figure}

We tested the stimulated jellyfish biohybrid robot without a forebody and with the addition of three different hemi-ellipsoid forebodies of increasing aspect ratio (see Methods). A total of $N=3$ different animals were tested with bell diameters $d_1=16.0 \pm0.5$ cm, $d_2=16.5 \pm0.5$ cm, and $d_3=18.0 \pm0.5$ cm. Each forebody configuration was tested 3 times in a randomized order to ensure that trends in forebody performance were not affected by any drift in animal performance over time. Sample biohybrid robot trajectories with and without the addition of a hemi-ellipsoid forebody $AR_3$ are shown in figure 5(a) and 5(b), respectively (see also Supplementary Movies 3 and 4). The images in figure 5 are equally spaced at intervals of 6.67 seconds and show a zoomed-in view of the middle one-third of the total height of the facility. An orange scaling dropline was marked every 1.5 meters to provide a reference length scale. The spatial intervals between images of the animal in figure 5(b) are larger than those in figure 5(a), qualitatively illustrating the faster swimming speed of the biohybrid robot with the addition of the streamlined mechanical forebody.

The biohybrid robot swimming speed was found to increase with the aspect ratio of the forebody, with the largest increase occurring between $AR_{0}$ (i.e., no forebody) and $AR_{1}$. Figure 5(c) plots the normalized biohybrid robot swimming speed (in body diameters per second) versus streamlining coefficient, $S$ (see Methods). Swimming speeds increased in direct proportion with the streamlining coefficient, reaching a plateau for the most streamlined forebodies. This plateau is likely due to the counteracting impacts of increased streamlining, inertia, and unsteady added mass. While larger forebodies have smaller $C_D$ values, they also displace significantly more mass and limit the ability of the biohybrid jellyfish robot to accelerate during the contraction phase. Diameter-normalized swimming speeds were observed to be larger for the larger jellyfish, likely reflecting a greater propulsive capacity to translate the mechanical forebodies downward against their positive buoyancy.

We also compared the swimming performance of each biohybrid robot to the same jellyfish without a mechanical forebody (figure 5(d)). The maximum enhancement factor (i.e., performance ratio relative to swimming without a forebody), which occurred for the mechanical forebody with the largest streamlining coefficient, i.e., $AR_{3}$, corresponded to an increase in swimming speeds of up to 61\%. Across all cases, we observed an average increase in swimming speed of $38\%\pm 22\%$. The performance plateau observed in figure 5(c) for the most streamlined forebodies persists here and the change in enhancement factor between forebodies $AR_2$ and $AR_3$ is not statistically significant. In light of prior work demonstrating an increase of up to 2.8 times the swimming speed of natural jellyfish \cite{xu2020low}, the present results suggest that the combined electromechanical enhancement could enable an overall increase in swimming speeds to more than 4.5 times natural, unstimulated jellyfish locomotion. Figure 5(c-d) was made using the MATLAB tracking program.

\begin{figure}[!h]
\includegraphics[width=1\textwidth]{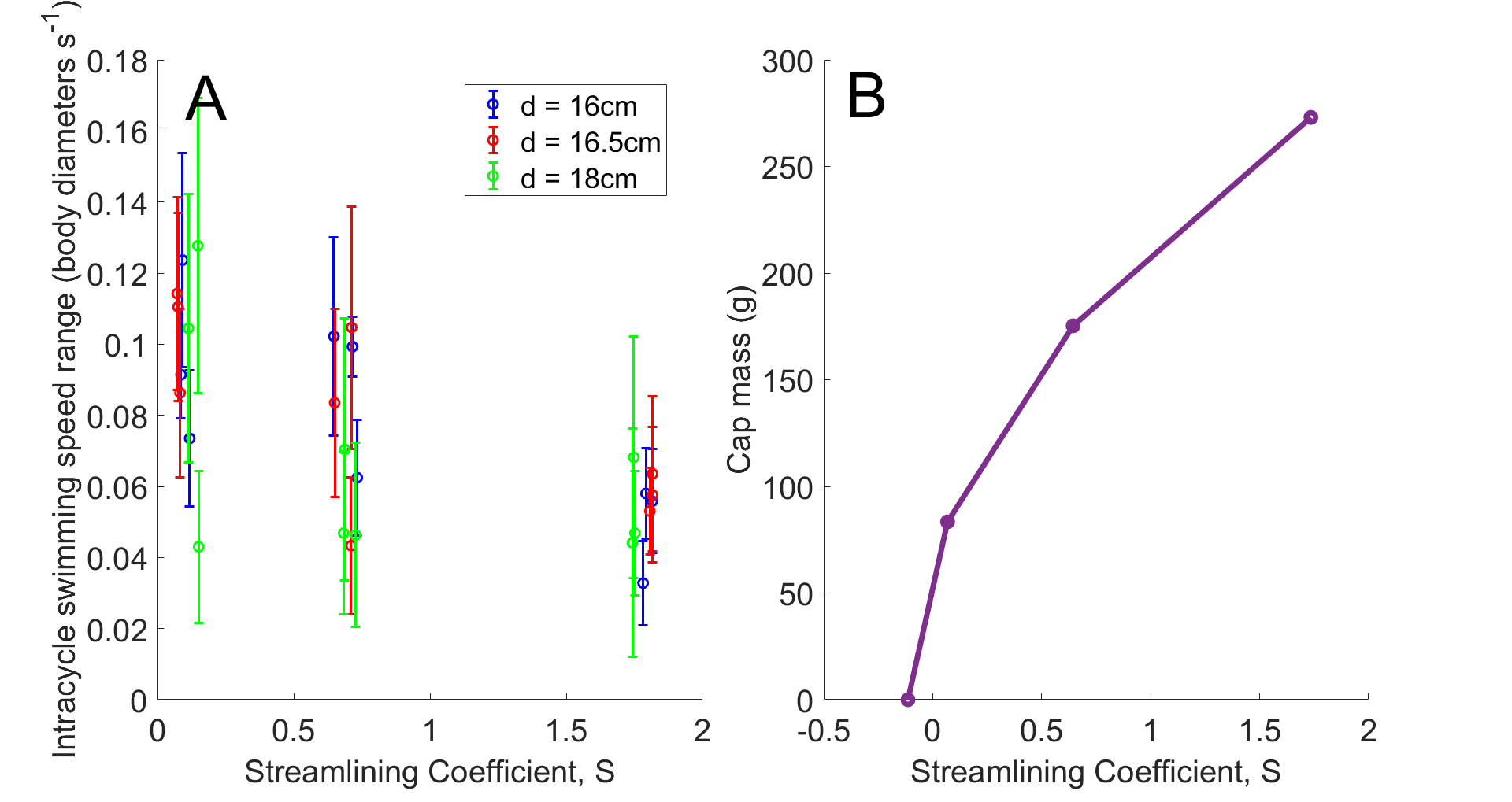}
\textbf{Figure 6. Intracycle swimming speed range and forebody inertia versus streamlining coefficient.} \\
(\textbf{A}) Intracycle swimming speed range decreases with streamlining coefficient $S$ due to increased inertia of the hemi-ellipsoid forebody. Error bars indicate $\pm$ 1 standard deviation of intracycle speed throughout each trial.  A small, random variability is added to the x-axis to help distinguish each of the data points. (\textbf{B}) Forebody mass increases with streamlining coefficient $S$. The largest tested forebody weighed $m_3=271.00 \pm .05$ grams in air.
\end{figure}

The addition of mechanical forebodies adds inertia to the biohybrid robotic system proportional to the volume of displaced water. The largest forebody corresponds to an increase in the mass by $m_3=271.04 \pm .05$ grams of payload and a volume of $105\% \pm 1\%$ of the animal body volume. Figure 6(a) plots the intracycle swimming speed range in body diameters per second vs $S$ of all 24 trials of the biohybrid robot with the three forebodies. With increasing streamlining coefficient $S$, the intracycle swimming speed range of the biohybrid robot decreased. Figure 6(b) illustrates that the forebody mass increased in direction proportion to the streamlining coefficient, which explains the concurrent reduction in intracycle swimming speed range. Specifically, because the biohybrid robot swam in a highly unsteady motion as shown in figure 6(a), the additional inertia changed the swimming dynamics. Figure 6(a) was made using the Blender tracking program.

\begin{figure}[!h]
\includegraphics[width=1\textwidth]{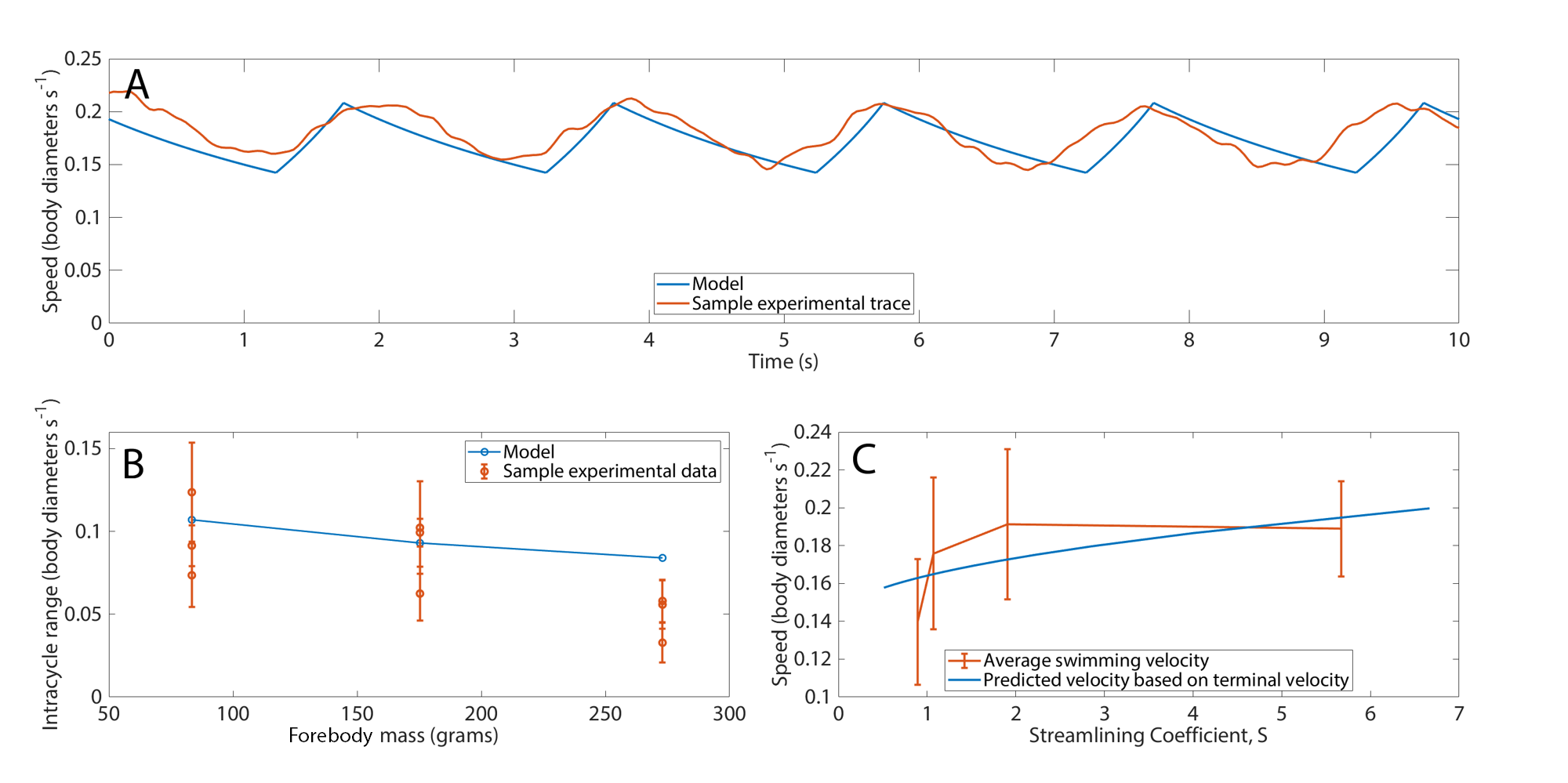}
\textbf{Figure 7. Validation of theoretical model of biohybrid robot performance.} \\
(\textbf{A}) Model prediction (blue) and empirical measurement (orange) of biohybrid robot swimmming speed for $AR_3$ forebody over 5 body contraction cycles. (\textbf{B}) Model prediction (blue) and empirical measurements (orange) of decrease in intracycle swimming speed range with increased forebody mass. Error bars indicate $\pm$ 1 standard deviation of the 3 trials at each point. (\textbf{C}) Terminal velocity plateau model prediction (blue) matches empirical measurements (orange) of average swimming speeds over all animals and all trials for each forebody configuration. Error bars indicate $\pm$ 1 standard deviation of all trials. 
\end{figure}

We developed a theoretical model of the biohybrid robot swimming dynamics to further investigate the dependence of the swimming performance on the forebody geometry and properties as well as the animal swimming behavior (see Methods). Figure 7(a) compares an example predicted swimming speed trace (blue) and measurements from swimming tests using the mechanical forebody $AR_{3}$ over 5 body contraction cycles. The model generally captures the trends in both swimming speed and intracycle speed range versus forebody mass (figure 7(b)). The overlayed data in figure 7(a-b) was made using the Blender tracking program. While the model predicted a decrease in the intracycle swimming speed range with increasing forebody mass, it underestimated the quantitative decrease observed between the $AR_2$ and $AR_3$ cases. We speculate that this might be due a change in the effective coefficient of drag of the combined biohybrid robotic system with the $AR_3$ forebody, given its smaller absolute drag coefficient.

Pooling the measurements from all 3 animals tested in these experiments, we observed a plateau in the average speed between the $AR_2$ and $AR_3$ cases. This plateau is qualitatively consistent with a scaled version of the terminal velocity equation, i.e., it indicates that performance at high swimming speeds is likely limited by the terminal velocity of the mechanical forebody. In sum, the model generally captures the biohybrid robot dynamics and may therefore be used to extrapolate to larger jellyfish and different forebody geometries in the future.

\section{\uppercase{Discussion}}
We have presented the electromechanical enhancement of a biohybrid jellyfish robot for ocean exploration through the addition of streamlined, passive forebodies. We demonstrated the ability to travel at speeds up to 4.5 times baseline jellyfish swimming speeds while carrying a payload of more than 270 grams in a volume of 105\% of the jellyfish volume. The flow around the biohybrid robot was examined using time-lapse pathline visualization to demonstrate local streamlining effects, and biohybrid robot swimming performance was investigated in a custom 6 m tall saltwater facility that enabled testing of swimming over distances significantly longer than prior studies. The addition of a payload was found to decrease intracycle acceleration, which may facilitate more precise onboard measurements in cases where unsteady platform motion should be minimized. By utilizing live animals and their ability to extract chemical energy from the water column via feeding, this approach has the potential to eliminate most of the battery power typically required for underwater propulsion, instead leveraging and controlling the jellyfish locomotion. This suggests an opportunity to expand the set of biohybrid robotic tools for ocean exploration.

While this work has focused on maximizing the swimming speed of the biohybrid jellyfish robots, it has not considered the associated metabolic costs of this increased performance. Previous work investigated the metabolic costs of natural jellyfish and found low muscle mass of less than 1\% of body mass resulting in low metabolic rates and a cost of transport lower than other animals \cite{gemmell2013passive}. Previous work on biohybrid jellyfish also found that increased swimming performance was not penalized by proportional increases in energy expenditure; in fact, the externally stimulated biohybrid jellyfish swam more efficiently than their natural counterparts \cite{xu2020low}. However, energy expenditure experiments have not been conducted for freely swimming biohybrid robotic jellyfish due to the challenges presented by the animals low-respiratory rates combined with large volumes of water required for free swimming resulting in imperceptible changes in oxygen concentration. Understanding the metabolic costs of the biohybrid jellyfish robot is important to developing a complete understanding of the increased performance. Future work will aim to address this knowledge gap using a combination of flow measurements and respirometry. Longer-term experiments on timescales relevant to ocean exploration (i.e., weeks to months in duration) will be necessary to confirm the sustainability of the performance observed here, while continuing to ensure the long-term well-being of the jellyfish. Finally, jellyfish in these experiments swam vertically to focus on vertical profiling applications in the ocean. Future work will extend the maneuverability of the biohybrid robotic jellyfish to enable three-dimensional exploration. 

\begin{figure}[!h]
\includegraphics[width=1\textwidth]{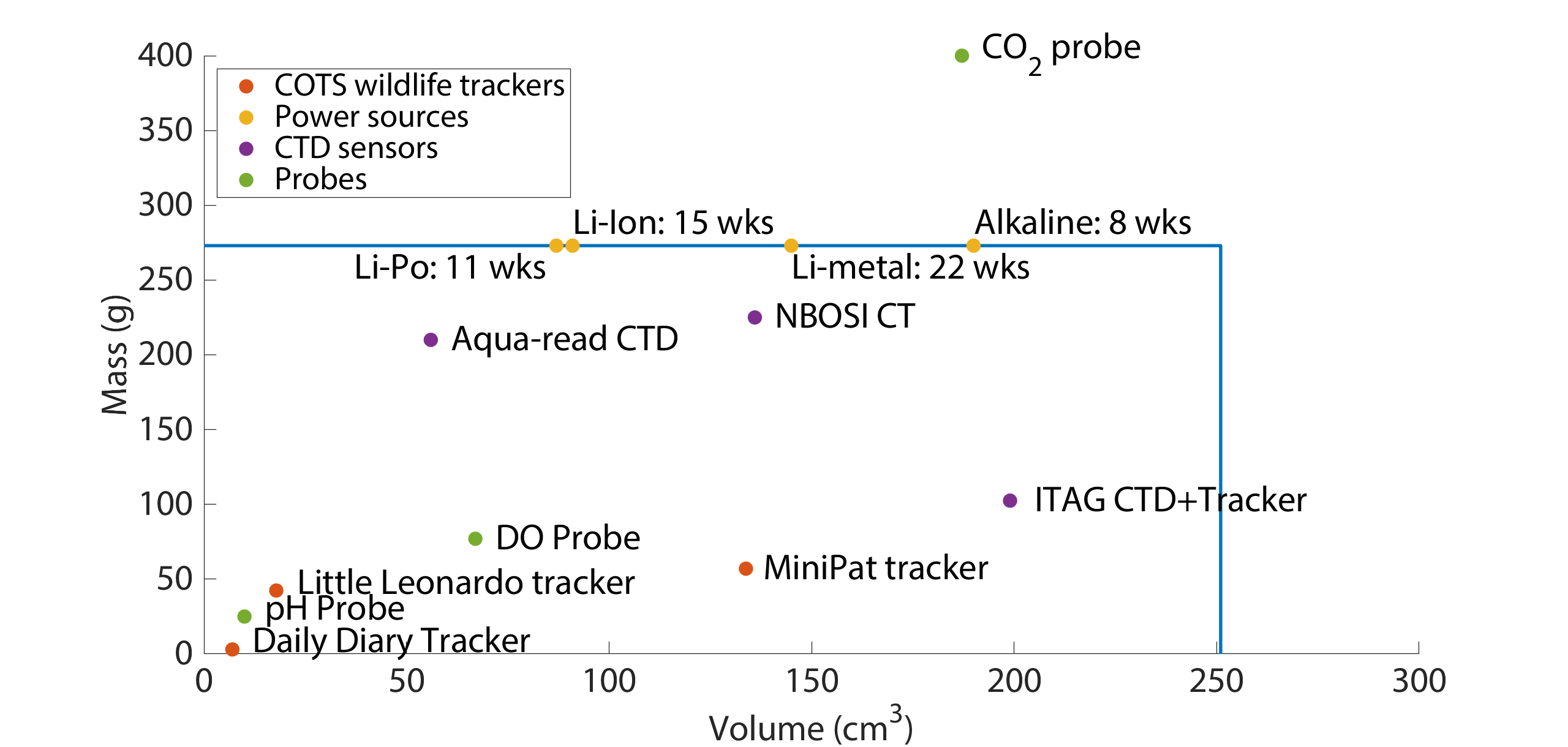}
\textbf{Figure 8. Utilizing biohybrid robotic jellyfish payload capacity for ocean science.} \\
Demonstrated payload mass and volume capacity (i.e., blue boundaries) can be utilized for a variety of ocean science and exploration applications. COTS wildlife trackers with a variety of sensors are shown in red; power sources with approximate duration powering current electronics in yellow; CTD sensors shown in purple; and chemical probes in green. Filling the forebody with a lithium metal battery could power the biohybrid robot for more than 22 weeks.
\end{figure}

The demonstrated ability to carry payloads could be utilized to investigate a variety of ocean science questions. Figure 8 explores some of the ways in which the mass and volume of the largest $AR_3$ hemi-ellipsoid forebody could be used to further ocean exploration. Several commercial off-the-shelf (COTS) wildlife trackers are shown in red, which include different combinations of sensors for parameters such as temperature, depth, light, and accelerometers for motion tracking (ORI400, Little Leonardo, Tokyo, Japan; MINIPAT, Wildlife Computers Inc, WA, USA; Daily Diary, Wildbyte Technologies Ltd, Swansea, United Kingdon). The forebody could instead be utilized as a large battery (power sources shown in orange) which could power the biohybrid robot for up to 22 weeks using a disposable 101.6 watt-hour Lithium-metal battery \cite{bradley2001power} or from 11 to 22 weeks using other battery types \cite{bradley2001power,manthiram2017outlook}. The relatively high energy density of lithium batteries could be exploited to preserve additional volume for less dense scientific payloads. For example, an onboard conductivity-temperature-depth (CTD) sensor, such as those shown in purple, could provide salinity and temperature measurements to inform ocean science models. Several COTS CTDs (LeveLine-MINI, AQUAREAD, Broadstairs, England; NBOSI CT, Neil Brown Ocean Sensors Inc., East Falmouth, MA, USA) could be accommodated within the forebody volume, as could a custom CTD with tracker from previous work \cite{mooney2015itag}. The biohybrid robot could also be used for ocean chemistry profiling (shown in green) by embedding a dissolved oxygen (DO) probe (Kit-103DX, Atlas Scientific, Long Island City, NY, USA), or to study ocean chemical plumes using a pH probe (ENV-20-pH, Atlas Scientific, Long Island City, NY, USA). While currently too large to fit into the forebody volumes tested in this work, a CO$_2$ probe to aid in the understanding of carbon distribution in the ocean may fit in a future forebody design (CO$_2$ probe, aquams, Maxeville, France). Deploying a combination of a larger battery and embedded sensors could power the biohybrid jellyfish robot for several weeks of exploration while recording using onboard sensors. Future work will continue to optimize this platform for longer deployment durations and additional scientific sensors. This versatility enables a wide variety of possible ocean science and exploration initiatives including improving ocean science models, investigating the concentrations of dissolved carbon dioxide, and studying the distribution of marine life in the ocean. 

Future work could also investigate jellyfish behavior while undergoing stimulation. This could include studying potential changes in feeding behavior with embedded electronics, as well as probing for changes after stimulation terminates, and studying tissue healing around embedded electrodes over long time-scales. There also exists an opportunity to study biohybrid robot jellyfish interactions with wild jellyfish through the field of animal-robot interactions \cite{romano2019review}.

This biohybrid robot itself can be fabricated for approximately \$20 plus the cost of the jellyfish, which can be procured from local aquaria or collected from the ocean. Fabrication of the robot utilizes consumer-grade 3D printers and manufacturing methods without specialized equipment. Production and deployment of these robots can be easily scaled. This accessibility of the technology can democratize ocean exploration and make it feasible to deploy swarms of biohybrid robots for comprehensive ocean exploration on timescales commensurate with a changing ocean.

\clearpage



\bibliographystyle{Science}

\begin{thebibliography}{10}

\bibitem{ramirez2010deep}
E.~Ramirez-Llodra, {\it et~al.\/}, {\it Biogeosciences\/} {\bf 7}, 2851 (2010).

\bibitem{levin2015deep}
L.~A. Levin, N.~Le~Bris, {\it Science\/} {\bf 350}, 766 (2015).

\bibitem{ryabinin2019decade}
V.~Ryabinin, {\it et~al.\/}, {\it Frontiers in Marine Science\/} {\bf 6}, 470 (2019).

\bibitem{wolfl2019seafloor}
A.-C. W{\"o}lfl, {\it et~al.\/}, {\it Frontiers in Marine Science\/} p. 283 (2019).

\bibitem{benway2019ocean}
H.~M. Benway, {\it et~al.\/}, {\it Frontiers in Marine Science\/} {\bf 6}, 393 (2019).

\bibitem{yuh2011applications}
J.~Yuh, G.~Marani, D.~R. Blidberg, {\it Intelligent service robotics\/} {\bf 4}, 221 (2011).

\bibitem{zereik2018challenges}
E.~Zereik, M.~Bibuli, N.~Mi{\v{s}}kovi{\'c}, P.~Ridao, A.~Pascoal, {\it Annual Reviews in Control\/} {\bf 46}, 350 (2018).

\bibitem{ventura2018mapping}
D.~Ventura, A.~Bonifazi, M.~F. Gravina, A.~Belluscio, G.~Ardizzone, {\it Remote Sensing\/} {\bf 10}, 1331 (2018).

\bibitem{reul2014sea}
N.~Reul, {\it et~al.\/}, {\it Surveys in Geophysics\/} {\bf 35}, 681 (2014).

\bibitem{parker1995marine}
D.~Parker, C.~Folland, M.~Jackson, {\it Climatic change\/} {\bf 31}, 559 (1995).

\bibitem{kavanaugh2021satellite}
M.~T. Kavanaugh, {\it et~al.\/}, {\it Oceanography\/} {\bf 34}, 62 (2021).

\bibitem{rudnick2016ocean}
D.~L. Rudnick, {\it Annual review of marine science\/} {\bf 8}, 519 (2016).

\bibitem{favali2006seafloor}
P.~Favali, L.~Beranzoli, {\it Annals of geophysic\/}  (2006).

\bibitem{sahoo2019advancements}
A.~Sahoo, S.~K. Dwivedy, P.~Robi, {\it Ocean Engineering\/} {\bf 181}, 145 (2019).

\bibitem{charette2010volume}
M.~A. Charette, W.~H. Smith, {\it Oceanography\/} {\bf 23}, 112 (2010).

\bibitem{katzschmann2018exploration}
R.~K. Katzschmann, J.~DelPreto, R.~MacCurdy, D.~Rus, {\it Science Robotics\/} {\bf 3}, eaar3449 (2018).

\bibitem{picardi2020bioinspired}
G.~Picardi, {\it et~al.\/}, {\it Science Robotics\/} {\bf 5}, eaaz1012 (2020).

\bibitem{li2021self}
G.~Li, {\it et~al.\/}, {\it Nature\/} {\bf 591}, 66 (2021).

\bibitem{carter2016navigating}
M.~I.~D. Carter, K.~A. Bennett, C.~B. Embling, P.~J. Hosegood, D.~J. Russell, {\it Movement Ecology\/} {\bf 4}, 1 (2016).

\bibitem{cuevas2008post}
E.~Cuevas, F.~A. Abreu-Grobois, V.~Guzm{\'a}n-Hern{\'a}ndez, M.~Liceaga-Correa, R.~P. Van~Dam, {\it Endangered Species Research\/} {\bf 10}, 123 (2008).

\bibitem{mooney2015itag}
T.~A. Mooney, {\it et~al.\/}, {\it Animal Biotelemetry\/} {\bf 3}, 1 (2015).

\bibitem{fossette2016tag}
S.~Fossette, {\it et~al.\/}, {\it Journal of Plankton Research\/} {\bf 38}, 1347 (2016).

\bibitem{chung2021review}
H.~Chung, J.~Lee, W.~Y. Lee, {\it Ocean Science Journal\/} {\bf 56}, 117 (2021).

\bibitem{sato2009remote}
H.~Sato, {\it et~al.\/}, {\it Frontiers in integrative neuroscience\/} {\bf 3}, 784 (2009).

\bibitem{sanchez2015locomotion}
C.~J. Sanchez, {\it et~al.\/}, {\it Journal of The Royal Society Interface\/} {\bf 12}, 20141363 (2015).

\bibitem{bozkurt2009balloon}
A.~Bozkurt, R.~F. Gilmour, A.~Lal, {\it IEEE transactions on biomedical engineering\/} {\bf 56}, 2304 (2009).

\bibitem{park2016phototactic}
S.-J. Park, {\it et~al.\/}, {\it Science\/} {\bf 353}, 158 (2016).

\bibitem{nawroth2012tissue}
J.~C. Nawroth, {\it et~al.\/}, {\it Nature biotechnology\/} {\bf 30}, 792 (2012).

\bibitem{schwefel2014wireless}
J.~Schwefel, {\it et~al.\/}, {\it Journal of The Electrochemical Society\/} {\bf 161}, H3113 (2014).

\bibitem{shoji2016autonomous}
K.~Shoji, K.~Morishima, Y.~Akiyama, N.~Nakamura, H.~Ohno, {\it 2016 IEEE International Conference on Mechatronics and Automation\/} (IEEE, 2016), pp. 629--634.

\bibitem{romano2023beehive}
D.~Romano, {\it Science Robotics\/} {\bf 8}, eadh1824 (2023).

\bibitem{purcell2012jellyfish}
J.~E. Purcell, {\it Annual review of marine science\/} {\bf 4}, 209 (2012).

\bibitem{purcell2010distribution}
J.~E. Purcell, R.~R. Hopcroft, K.~N. Kosobokova, T.~E. Whitledge, {\it Deep Sea Research Part II: Topical Studies in Oceanography\/} {\bf 57}, 127 (2010).

\bibitem{gallo2015submersible}
N.~D. Gallo, {\it et~al.\/}, {\it Deep Sea Research Part I: Oceanographic Research Papers\/} {\bf 99}, 119 (2015).

\bibitem{purcell2007environmental}
J.~E. Purcell, {\it Marine Ecology Progress Series\/} {\bf 348}, 183 (2007).

\bibitem{gemmell2013passive}
B.~J. Gemmell, {\it et~al.\/}, {\it Proceedings of the National Academy of Sciences\/} {\bf 110}, 17904 (2013).

\bibitem{xu2020low}
N.~W. Xu, J.~O. Dabiri, {\it Science Advances\/} {\bf 6}, eaaz3194 (2020).

\bibitem{xu2020ethics}
N.~Xu, O.~Lenczewska, S.~Wieten, C.~Federico, J.~Dabiri  (2020).

\bibitem{Xu2024transanimalism}
N.~Xu, J.~Dabiri, {\it The creation of an augmented jellyfish: Ethical considerations from a scientific perspective\/} (2024).

\bibitem{smith2009nociceptors}
E.~S.~J. Smith, G.~R. Lewin, {\it Journal of Comparative Physiology A\/} {\bf 195}, 1089 (2009).

\bibitem{xu2020field}
N.~W. Xu, {\it et~al.\/}, {\it Biomimetics\/} {\bf 5}, 64 (2020).

\bibitem{batchelor1967introduction}
G.~K. Batchelor, {\it An introduction to fluid dynamics\/} (Cambridge university press, 1967).

\bibitem{hoerner1958fluid}
S.~F. Hoerner, {\it Fluid-dynamic drag\/} (1958).

\bibitem{daniel1983mechanics}
T.~L. Daniel, {\it Canadian Journal of Zoology\/} {\bf 61}, 1406 (1983).

\bibitem{mchenry2003ontogenetic}
M.~J. McHenry, J.~Jed, {\it Journal of experimental biology\/} {\bf 206}, 4125 (2003).

\bibitem{bradley2001power}
A.~M. Bradley, M.~D. Feezor, H.~Singh, F.~Y. Sorrell, {\it IEEE Journal of oceanic Engineering\/} {\bf 26}, 526 (2001).

\bibitem{manthiram2017outlook}
A.~Manthiram, {\it ACS central science\/} {\bf 3}, 1063 (2017).

\bibitem{romano2019review}
D.~Romano, E.~Donati, G.~Benelli, C.~Stefanini, {\it Biological cybernetics\/} {\bf 113}, 201 (2019).

\end{thebibliography}
\nobibliography{scibib}

\section*{References}

\begin{enumerate}
    
\item E. Ramirez-Llodra, A. Brandt, R. Danovaro, B. De Mol, E. Escobar, C. R. German, L. A. Levin, P. Martinez Arbizu, L. Menot, P. Buhl-Mortensen, B. E. Narayanaswamy, C. R. Smith, D. P. Tittensor, P. A. Tyler, A. Vanreusel, and M. Vecchione, Deep, diverse and definitely different: unique attributes of the world's largest ecosystem. \textit{Biogeosciences} 7, 2851 (2010).
\item L. A. Levin, N. Le Bris, The deep ocean under climate change. Science 350, 766 (2015).
\item V. Ryabinin1, J. Barbière, P. Haugan, G. Kullenberg, N. Smith, C. McLean, A. Troisi, A. Fischer, S. Aricò, T. Aarup, P. Pissierssens, M.Visbeck, H. O. Enevoldsen, J. Rigaud, The UN decade of ocean science for sustainable development. \textit{Frontiers in Marine Science} 6, 470 (2019).
\item A. C. Wölfl, H. Snaith, S. Amirebrahimi, C. W. Devey, B. Dorschel, V. Ferrini, V. A. I. Huvenne, M. Jakobsson, J. Jencks, G. Johnston, G. Lamarche, L. Mayer, D. Millar, T. Haga Pedersen, K. Picard, A. Reitz, T. Schmitt, M. Visbeck, P. Weatherall, R. Wigley, Seafloor mapping--the challenge of a truly global ocean bathymetry. \textit{Frontiers in Marine Science} p. 283 (2019).
\item H. M. Benway, L. Lorenzoni, A. E. White, B. Fiedler, N. M. Levine, D. P. Nicholson, M. D. DeGrandpre, H. M. Sosik, M. J. Church, T. D. O’Brien, M. Leinen, R. A. Weller, D. M. Karl, S. A. Henson1, R. M. Letelier, Ocean time series observations of changing marine ecosystems: an era of integration, synthesis, and societal applications. \textit{Frontiers in Marine Science} 6, 393 (2019).
\item J. Yuh, G. Marani, D. R. Blidberg, Applications of marine robotic vehicles. \textit{Intelligent service robotics} 4, 221 (2011).
\item E. Zereik, M. Bibuli, N. Miskovi C., P. Ridao, A. Pascoal, Challenges and future trends in marine robotics. \textit{Annual Reviews in Control} 46, 350 (2018).
\item D. Ventura, A. Bonifazi, M. F. Gravina, A. Belluscio, G. Ardizzone, Mapping and classification of ecologically sensitive marine habitats using unmanned aerial vehicle (UAV) imagery and object-based image analysis (OBIA). \textit{Remote Sensing} 10, 1331 (2018).
\item N. Reul, S. Fournier, J. Boutin, O. Hernandez, C. Maes, B. Chapron, G. Alory, Y. Quilfen, J. Tenerelli, S. Morisset, Y. Kerr, S. Mecklenburg and S. Delwart, Sea surface salinity observations from space with the SMOS satellite: A new means to monitor the marine branch of the water cycle. \textit{Surveys in Geophysics} 35, 681 (2014).
\item D. Parker, C. Folland, M. Jackson, Marine surface temperature: observed variations and data requirements. \textit{Climatic change} 31, 559 (1995).
\item M. T. Kavanaugh, T. Bell, D. Catlett, M. A. Cimino, Scott C. Doney, Willem Klajbor, Monique Messié, Enrique Montes, Frank E. Muller-Karger, Daniel Otis, Jarrod A. Santora, Isaac D. Schroeder, Joaquin Triñanes, David A. Siegel, Satellite remote sensing and the marine biodiversity observation network. \textit{Oceanography} 34, 62 (2021).
\item D. L. Rudnick, Ocean research enabled by underwater gliders. \textit{Annual review of marine science} 8, 519 (2016).
\item P. Favali, L. Beranzoli, Seafloor observatory science: A review. Annals of geophysic (2006).
\item A. Sahoo, S. K. Dwivedy, P. Robi, Advancements in the field of autonomous underwater vehicle. \textit{Ocean Engineering 181}, 145 (2019).
\item M. A. Charette, W. H. Smith, The volume of Earth's ocean. Oceanography 23, 112 (2010).
\item R. K. Katzschmann, J. DelPreto, R. MacCurdy, D. Rus, Exploration of underwater life with an acoustically controlled soft robotic fish. \textit{Science Robotics} 3, eaar3449 (2018).
\item G. Picardi, M. Chellapurath, S. Iacoponi S. Stefanni, C. Laschi, M. Calisti, Bioinspired underwater legged robot for seabed exploration with low environmental disturbance. \textit{Science Robotics} 5, eaaz1012 (2020).
\item G. Li, X. Chen, F. Zhou, Y. Liang, Y. Xiao, X. Cao, Z. Zhang, M. Zhang, B. Wu, S. Yin, Y. Xu, H. Fan, Z. Chen, W. Song, W. Yang, B. Pan, J. Hou, W. Zou, S. He, X. Yang, G. Mao, Z. Jia, H. Zhou, T. Li, S. Qu, Z. Xu, Z. Huang, Y. Luo, T. Xie, J. Gu, S. Zhu, W. Yang, Self-powered soft robot in the Mariana Trench. \textit{Nature} 591, 66 (2021).
\item M. I. D. Carter, K. A. Bennett, C. B. Embling, P. J. Hosegood, D. J. Russell, Navigating uncertain waters: a critical review of inferring foraging behavior from location and dive data in pinnipeds. \textit{Movement Ecology} 4, 1 (2016).
\item E. Cuevas, F. A. Abreu-Grobois, V. Guzman-Hernandez, M. Liceaga-Correa, R. P. Van Dam, Post-nesting migratory movements of hawksbill turtles Eretmochelys imbricata in waters adjacent to the Yucatan Peninsula, Mexico. \textit{Endangered Species Research} 10, 123 (2008).
\item T. A. Mooney, K. Katija, K. A. Shorter, T. Hurst, J. Fontes, P. Afonso, ITAG: an eco-sensor for fine-scale behavioral measurements of soft-bodied marine invertebrates. \textit{Animal Biotelemetry} 3, 1 (2015).
\item S. Fossette, K. Katija, J. A. Goldbogen, S. Bograd, W. Patry, M. J. Howard, T. Knowles, S. H.D. Haddock, L. Bedell, E. L. Hazen, B. H. Robison, T. A. Mooney, K. A. Shorter, T. Bastian, A. C. Gleiss, How to tag a jellyfish? A methodological review and guidelines to successful jellyfish tagging. \textit{Journal of Plankton Research} 38, 1347 (2016).
\item H. Chung, J. Lee, W. Y. Lee, A review: Marine bio-logging of animal behaviour and ocean environments. \textit{Ocean Science Journal} 56, 117 (2021).
\item H. Sato, C. W. Berry, Y. Peeri, E. Baghoomian, B. E. Casey, G. Lavella, J. M. VandenBrooks, J. F. Harrison, M. M. Maharbiz, Remote radio control of insect flight. \textit{Frontiers in Integrative Neuroscience} 3, (2009).

\item C. J. Sanchez, C.-W. Chiu, Y. Zhou, J. M. González, S. B. Vinson, H. Liang, Locomotion control of hybrid cockroach robots. \textit{Journal of The Royal Society Interface} 12, 20141363 (2015).
\item A. Bozkurt, R. F. Gilmour, A. Lal, Balloon-assisted flight of radio-controlled insect biobots. \textit{IEEE transactions on biomedical engineering} 56, 2304 (2009).
\item S.-J. park, M. Gazzola, K. S. Park, S. Park, V. D. Santo, E. L. Blevins, J. U. Lind, P. H. Campbell, S Dauth, A K. Capulli, F S. Pasqualini, S Ahn, A Cho, H. Yuan, B. M. Maoz, R. Vijaykumar, J.-W. Choi, K Deisseroth, G V. Lauder, L. Mahadevan, K. K. Parker, Phototactic guidance of a tissue-engineered soft-robotic ray. \textit{Science} 353, 158 (2016).
\item J. C. Nawroth, H. Lee, A.W. Feinberg, C. M. Ripplinger, M. L. McCain, A. Grosberg, J. O. Dabiri, K. K. Parker, A tissue-engineered jellyfish with biomimetic propulsion. \textit{Nature biotechnology} 30, 792 (2012).

\item J. Schwefel, R. E. Ritzmann, I. N. Lee, A. Pollack, W. Weeman, S. Garverick, M. Willis, M. Rasmussen, and D. Scherson, Wireless Communication by an Autonomous Self-Powered Cyborg Insect. \textit{Journal of The Electrochemical Society} 161, H3113, (2015).

\item K. Shoji, K. Morishima, Y. Akiyama, N. Nakamura, H. Ohno, Autonomous environmental monitoring by self-powered biohybrid robot, \textit{IEEE International Conference on Mechatronics and Automation}, (2016).

\item D. Romano, The beehive of the future is a robot socially interacting with honeybees. \textit{Science Robotics}, 8, 76, (2023).

\item J. E. Purcell, Jellyfish and ctenophore blooms coincide with human proliferations and environmental perturbations. \textit{Annual review of marine science} 4, 209 (2012).
\item J. E. Purcell, R. R. Hopcroft, K. N. Kosobokova, T. E. Whitledge, Distribution, abundance, and predation effects of epipelagic ctenophores and jellyfish in the western Arctic Ocean. \textit{Deep Sea Research Part II: Topical Studies in Oceanography} 57, 127 (2010).
\item N. D. Gallo, J. Cameron, K. Hardy, P. Fryer, D. H. Bartlett, L. A. Levin, Submersible-and lander-observed community patterns in the Mariana and New Britain trenches: influence of productivity and depth on epibenthic and scavenging communities. \textit{Deep Sea Research Part I: Oceanographic Research Papers} 99, 119 (2015).
\item J. E. Purcell, Environmental effects on asexual reproduction rates of the scyphozoan Aurelia labiate. \textit{Marine Ecology Progress Series} 348, 183 (2007).
\item B. J. Gemmell, J. H. Costello, S. P. Colin, J. O. Dabiri, D Tafti, S Priya, Passive energy recapture in jellyfish contributes to propulsive advantage over other metazoans. \textit{Proceedings of the National Academy of Sciences} 110, 17904 (2013).
\item N. W. Xu, J. O. Dabiri, Low-power microelectronics embedded in live jellyfish enhance propulsion. \textit{Science Advances} 6, eaaz3194 (2020).
\item N. Xu, O. Lenczewska, S. Wieten, C. Federico, J. Dabiri, Ethics of Biohybrid Robotic Jellyfish Modification and Invertebrate Research. \textit{Preprints} 2020, 2020100008.
\item N. W. Xu, J. O. Dabiri, The creation of an augmented jellyfish: Ethical considerations from a scientific perspective. \textit{Transanimalism}, 8, 127-139 (2024).
\item E. S. J. Smith, G. R. Lewin, Nociceptors: a phylogenetic view. \textit{Journal of Comparative Physiology} A 195, 1089 (2009).
\item N. W. Xu, J. P. Townsend, J. H. Costello, S. P. Colin, B. J. Gemmell, J. O. Dabiri, Field testing of biohybrid robotic jellyfish to demonstrate enhanced swimming speeds. \textit{Biomimetics} 5, 64 (2020).

\item G. K. Batchelor, \textit{An Introduction to Fluid Dynamic}s (Cambridge university press, 1967).
\item S. F. Hoerner, \textit{Fluid-Dynamic Drag}. (Bricktown, NJ, 1958).
\item T. L. Daniel, Mechanics and energetics of medusan jet propulsion. \textit{Canadian Journal of Zoology} 61, 1406 (1983).
\item M. J. McHenry, J. Jed, The ontogenetic scaling of hydrodynamics and swimming performance in jellyfish (Aurelia aurita). \textit{Journal of experimental biology} 206, 4125 (2003).

\item A. M. Bradley, M. D. Feezor, H. Singh, F. Y. Sorrell, Power systems for autonomous underwater vehicles. \textit{IEEE Journal of oceanic Engineering} 26, 526 (2001).
\item A. Manthiram, An outlook on lithium ion battery technology. \textit{ACS central science} 3, 1063 (2017).
\item D. Romano, E. Donati, G. Benelli, C. Stefanini, A review on animal–robot interaction: from bio-hybrid organisms to mixed societies. \textit{Biological cybernetics} 113 (2019).

\end{enumerate}

\section*{Acknowledgments}
The authors would like to thank Cabrillo Marine Aquarium for providing the jellyfish used in these experiments. \textbf{Funding:} This work was supported by the National Science Foundation Alan T. Waterman Award and NSF Graduate Research Fellowship grant number DGE‐1745301. \textbf{Author contributions:} S.R.A. and J.O.D. conceived of project, S.R.A. conducted experiments, S.R.A. and J.O.D. analyzed results and wrote paper. \textbf{Competing interests:} The authors declare that they have no competing financial interests. \textbf{Data and materials availability:} Contact J.O.D. for data and source code.


\section*{Supplementary materials}
Video S1. Sample video of free swimming biohybrid jellyfish robot swimming through tracer particles without attached forebody. Video is slowed down at 0.5 times speed.\\
Video S2. Sample video of free swimming biohybrid jellyfish robot swimming through tracer particles with attached $AR_3$ forebody. Video is slowed down at 0.5 times speed.\\
Video S3. Sample timelapse video of free swimming biohybrid jellyfish robot swimming in the 6 m saltwater facility without attached forebody. Video is sped up at 15 times speed.\\
Video S4. Sample timelapse video of free swimming biohybrid jellyfish robot swimming in the 6 m saltwater facility with attached $AR_3$ forebody. Video is sped up at 15 times speed.\\


\clearpage

\end{document}